\documentclass[preprint]{aastex}

\begin{document}

\title{Deep HST+STIS Color-Magnitude Diagrams of the Dwarf Irregular Galaxy WLM:
Detection of the Horizontal Branch
\footnote{Based on observations obtained with the NASA-ESA Hubble Space Telescope, and with the ESO NTT telescope at La Silla Observatory.}}

\author{Marina Rejkuba $^{2,3}$, Dante Minniti $^{3,4}$,
Michael D. Gregg$^{4}$, Albert A. Zijlstra $^{5}$, M. Victoria Alonso$^{6}$, 
Paul Goudfrooij$^{7,8}$}

\altaffiltext{2}{European Southern Observatory, Karl-Schwarzschild-Strasse
2, D-85748 Garching, Germany\\
E-mail: mrejkuba@eso.org}

\altaffiltext{3}{Department of Astronomy, P. Universidad Cat\'olica, Casilla
306, Santiago 22, Chile\\
E-mail: mrejkuba@astro.puc.cl, dante@astro.puc.cl}

\altaffiltext{4}{Lawrence Livermore National Laboratory, MS L-413, P.O. Box 808,
Livermore, CA 94550}

\altaffiltext{5}{Department of Physics, UMIST, P.O.Box 88, 
Manchester M60 1QD, UK\\
E-mail: a.zijlstra@umist.ac.uk}

\altaffiltext{6}{Observatorio Astron\'omico de C\'ordoba, Laprida 854, 
C\'ordoba, Argentina\\
E-mail: vicky@oac.uncor.edu}

\altaffiltext{7}{Space Telescope Science Institute, 3700 San Martin Drive,
Baltimore, MD 21218, USA\\
E-mail: goudfroo@stsci.edu}

\altaffiltext{8}{Affiliated with the Astrophysics Division, Space Science
Department, European Space Agency}


\begin{abstract}
We have obtained V and I-band photometry for 1886 stars down to $I=27$
and $V=28$ in the field of the dwarf irregular galaxy WLM, using deep
STIS $CL$ and $LP$-band images taken in the parallel mode with the
Hubble Space Telescope. The photometry is used to study the horizontal
branch identified in WLM. The horizontal branch, extending blueward
from the red giant clump, is an unambiguous signature of an old
population.  We demonstrate that it is possible to reach the
horizontal branch of an old population at a distance of $\simeq 1$ Mpc
using STIS, with relatively short exposure times.  {}From the VI
color--magnitude diagram we obtain an accurate distance modulus
$(m-M)_0 = 24.95 \pm 0.13$ for WLM by using the V-magnitude of the
horizontal branch, and by adopting $E(V-I) = 0.03$.  The implications
are: 1) WLM formed stars at high redshift, and 2) the old population of WLM
can be representative of a protogalactic fragment, related to those accreted
to form the Milky Way halo.
\end{abstract}

\keywords{Galaxies: individual (WLM, DDO 221, A2359-1544, UGC A444) --
Galaxies: irregular -- Galaxies: stellar content -- Local Group}

\section{Introduction}

In 1997 February, the second Space Shuttle servicing mission to the Hubble
Space Telescope (HST) 
installed the Space Telescope Imaging Spectrograph (STIS). This
new spectrograph can be used in direct imaging mode providing very deep
high-resolution images. The two optical wideband choices on STIS are a clear
aperture ($CL$) and a longpass ($\lambda > $5500 \AA) filter ($LP$). Based on
these two wide bands, Gregg \& Minniti (1997) defined an effective shortpass 
filter (SP), thus establishing a two color photometric system. They
suggested that these unusually wide bandpasses of STIS offer the possibility
to reach much fainter limiting magnitudes with respect to the 
other HST imaging instruments like the Wide Field
Planetary Camera 2 (WFPC2). STIS is expected to reach 
$\sim$1.5 mag fainter than WFPC2 
for a given integration time. This advantage of STIS over WFPC2 is due 
to three factors: (i) much broader passbands; (ii) the higher quantum 
efficiency and (iii) the lower readout noise of STIS's CCD. 

By using spectrophotometric stellar libraries, Gregg \& Minniti
(1997) have argued that it is possible to 
convert the STIS magnitudes and (SP$-$LP)
color to the standard photometric systems. They derived preliminary 
transformations to the VI Kron--Cousins wideband system. These transformations
were based upon preflight specifications of the instrument and had to be
empirically calibrated. Such an empirical calibration is now possible for
the data collected during 
the STIS Parallel Survey (see, e.g., Gardner et al.\ 1998).

The STIS Parallel Survey  data contain observations of the 
dwarf irregular galaxy DDO221 (= WLM)
located at the edge of the Local Group (Sandage \& Carlson 1985; van den
Bergh 1994). In a photometric study of this galaxy, Minniti \& Zijlstra
(1996, 1997) presented VI photometry obtained with the NTT at La Silla
Observatory. There is an overlap between he STIS field of WLM
and the NTT field, enabling us to establish the empirical
calibrations of the STIS photometric system.
Based on this calibration, we construct a VI color--magnitude diagram (CMD)
that is much deeper than those of any other Local Group galaxy studied so far
with a comparable amount of exposure time (see Section~2.2), reaching
V$\sim$28 and  I$\sim$27.
The analysis of the CMD provides us with the 
fundamental properties of WLM's
stellar populations, such as, metallicities, reddening, age and distance
(Lee 1993; Minniti and Zijlstra 1996; Dolphin 2000). 
These properties are essential for
establishing the formation history and evolution of WLM, a typical dwarf
irregular galaxy that is representative of the most numerous group of 
galaxies in the
Universe (Dekel \& Silk 1986). The importance of WLM is stressed by the fact
that it is located far from any other large spiral member of the Local
Group, and therefore most probably has remained isolated over the Hubble time
(Peebles 1995). The nearest galaxy to WLM is the recently discovered 
Cetus dSph, at a distance of $\sim175$ kpc (Whiting et al.\ 1999). Having
remained isolated, WLM carries the unperturbed imprint of its formation
and evolution in its stellar content. 

The observations, data reductions, and photometry of WLM, and the
resulting color--magnitude diagrams are given in Section 2. An
empirical calibration of the STIS photometric system is given in
section 2.3.  Fundamental parameters of WLM (metallicity, age and
distance) are redetermined in Section 3.  The conclusions and
implications of this work are summarized in Section 4.

\section {The Data}

\subsection {Ground-based Photometry}

The ground-based 
observations of WLM were obtained during the night of September 5, 1995,
as part of a long term program monitoring the variable stars
in several galaxies (Zijlstra, Minniti \& Brewer 1997). 
We used the red arm (RILD mode)
of the ESO Multi-Mode Instrument (EMMI) on the New Technology
Telescope (NTT) at La Silla Observatory, operated remotely from the
ESO headquarters in Garching, Germany.  
We used the 2048$\times$2048 Tek
CCD with a scale of $0.268\arcsec$ pix$^{-1}$.  The weather was 
photometric, with seeing varying from $0.5\arcsec$ to $1.0\arcsec$. 

The observations of WLM were secured with the V and Cousins I 
filters. They consist of pairs of frames with exposure times of 900 sec
in I (airmass 1.1), and 1800 sec in V (airmass 1.1). These frames, 
covering an area of $9\arcmin \times 9\arcmin$, are
centered on the galaxy.
Figure~1 shows a portion, covering only $1\arcmin \times 1\arcmin$,  
of the NTT+EMMI images.

The reductions of the CCD frames were carried out by following standard
procedures, using the package CCDRED within the IRAF environment. 
The pairs of images observed through each filter were combined. 
The final FWHM of stellar images were 0.8 and 0.7 arcsec in V and I 
respectively.

A total of 5 standards 
were observed during the night in order to calibrate our photometry.
The photometric transformations to the standard system
were done following the procedure described
by Minniti \& Zijlstra (1997).
These transformations reproduce, with an rms of 0.016 in V and 0.031 in I, 
the magnitudes of the standard stars.

The photometric measurements were performed within the IRAF
environment using DAOPHOT II, an improved version of the
original DAOPHOT package developed by Stetson (1987).  In particular,
DAOPHOT II can accommodate a spatially varying point-spread function
across the field. All stars in the field of WLM that were more 
than 4$\sigma$ above
the background in the V and I frames were located, and their magnitudes
measured by fitting a Moffat point-spread function. The resulting
NTT limiting magnitudes ($4\sigma$) were $V= 24$, and $I= 23.5$.

\subsection {STIS Photometry}

The STIS images were acquired through the HST parallel program 7912
in September 1998.
The center of the observed field was 
$\alpha_{2000}=00^{\rm h} 02^{\rm m} 03.19^{\rm s}$, 
$\delta_{2000}=-15^\circ 23\arcmin 57.5\arcsec$, at the N-E edge of
WLM's disk.
The $LP$-band image covers $28\arcsec \times 50\arcsec$, and
the $CL$-band image covers $50\arcsec \times 50\arcsec$, 
but only the overlap region has been used.
Four images with a total exposure time of 1680 seconds were combined to
obtain the final $CL$-band image, 
and four images with a total exposure time of 
2665 seconds were combined to
obtain the final $LP$-band image. 
Cosmic rays were identified and eliminated in
this process. 

In the $28\arcsec \times 50\arcsec$ STIS images we detect 1886 point 
sources, reaching more than 3 magnitudes fainter than the faintest stars 
seen in the NTT images. In particular,
some of the single sources in the NTT images are resolved into multiple
sources by STIS. Objects containing bad pixels were eliminated from our final
photometry list.

Figure~2 shows the STIS $LP$-band image, with the orientation indicated in
Figure~1. This parallel observation with STIS falls at the edge of the WLM 
disk, and is expected to include disk and halo stars.

The photometry was performed with DAOPHOT~II package in IRAF. For the
zero-point calibration we followed instructions in the 
``HST Data Handbook'' (1997). Using the values of the 
inverse sensitivity (PHOTFLAM) and HST zero-point (PHOTZPT) keywords in the
headers of the images, zero-points of 26.067 and 26.518 were calculated for
$LP$ and $CL$ respectively. The aperture corrections applied to the magnitudes
measured through apertures of 2 pixels in diameter were $-0.575$ for $LP$ 
and $-0.484$ for $CL$.

The STIS CCD suffers from imperfect Charge Transfer Efficiency (CTE),
as recently reported by Gilliland, Goudfrooij \& Kimble (1999). We
have used their formulation of the CTE to estimate its effect on our
photometry. It turns out that, except for very faint stars that are
near our detection threshold ($V \ga 27.5$) and near the bottom of the
CCD, the effect on our magnitudes is below 0.01 mag, which is
negligible for our purposes.  Figure~3 shows the photometric errors as
function of instrumental $CL$ and $LP$ magnitudes.

We note that
since the galaxy/star ratio increases rapidly at faint magnitudes (Tyson 1988),
the background galaxy contamination also has to be  taken into account.
Using the number counts in the STIS field of the Hubble Deep Field South image
(Gardner et al.\ 2000), we estimate that in the field $\sim$85 galaxies 
are present 
between $CL$ magnitudes of 26 and 28. Most of the galaxies are 
resolved on our STIS images 
and are discarded by our choice of the ``sharpness'' parameter 
in the DAOPHOT~II routines.
Only a few fainter and very compact galaxies may be confused with stars, but
these do not have a significant impact on the present work. 

Figure 4 shows the $CL ~vs ~CL-LP$, and $LP ~vs ~CL-LP$ instrumental
color-magnitude diagrams for a total of 1886 stars
with centroids matched in $CL$ and $LP$ frames to better than 1.0 pixels
(0\farcs05), and which satisfy stringent criteria of photometric quality
($\sigma \leq 0.5$, $\chi \leq 2$, and sharpness $\geq -1$). The
sharpness criterion eliminates galaxies as well as possible star clusters at
the distance of WLM (see Sect.~3.1).  
The filled circles in Figure~4 
represent the stars in common with the NTT images (41 in
total) that are used for the empirical calibration of the STIS photometry.

\subsection {Empirical Calibration of the STIS Photometric System}

After STIS became operational in space, a significantly lower throughput
of the CCD+LP system has been observed (e.g. Gardner et al. 2000). 
Although it is still possible to define
the effective short bandpass by subtracting, from the $CL$ flux, the $LP$ flux 
scaled by 1.31, we use here only the $CL$ and $LP$
bandpasses for calibration.  
This simplifies the estimation of the uncertainties in the final photometry.

The photometry of the 41 STIS stars in common with the NTT observations is
listed in Table~1. Respectively, columns 2 and 3 give
the x and y positions of the stars on the NTT chip, 
columns 4 and 5 give V and I magnitudes of these stars, 
columns 6, 7, 8 and 9 give x and y positions 
of the stars on the STIS chip and their $CL$ and $LP$ magnitudes.
For some of these stars careful inspection of the STIS images showed crowding, 
i.e. more than 2-3 stars with different
magnitudes at the position of one NTT ``star''. These were not used for
the calibration and are flagged with two asterisks 
in the last column of Table~1. Two
more calibrators were rejected because of their marginal detection in the NTT
image which carries a big magnitude error (flagged with one 
asterisk in Table~1). 

The remaining 33 stars were used to transform $CL$ into V magnitude, 
and $LP$ into I
magnitude. Figure~5 shows the comparison of the photometry.
The relationships are approximately linear, as expected within
the color range  $-0.8<V-I<2.0$ under consideration (Gregg \& Minniti 1997).
Using a least square fit the following transformation equations were
obtained:

\begin{equation}
(V-I) = 1.76 (\pm 0.13) \times (CL-LP) + 1.18 (\pm 0.03) 
\end{equation}
\begin{equation}
(V-CL) = 0.26 (\pm 0.19) \times (CL-LP) +  0.01 (\pm 0.04)
\end{equation}
\begin{equation}
(I-LP) = -0.51 (\pm 0.14) \times (CL-LP) - 1.17  (\pm 0.03)
\end{equation}

The scatter about the mean relations is significant: $\sigma_{\rm
rms}=0.16$, 0.24 and 0.17 for the first, second and third equation
respectively.  This is largely due to the NTT photometric errors at
faint magnitudes ($\sigma_V>0.08$ for $V>22.5$, $\sigma_I>0.10$ for
$I>22$; Minniti \& Zijlstra 1997), and the small number of acceptable
calibrators in the blue part of the color range,
$-0.8<V-I<-0.4$. There is also an inherent scatter because of the
differences in widths and central wavelengths between $CL$
($\lambda_c$=5851.5\AA) and V ($\lambda_c$=5426\AA) as well as between
$LP$ ($\lambda_c$=7228.5\AA) and I ($\lambda_c$=7985\AA) filters.
This scatter will depend on abundances, temperature and surface
gravities, but because of the homogeneity of our comparison stars,almost all
being red giants or supergiants, we estimate that it is smaller than the
scatter induced by the NTT photometric errors. The complete photometry list,
given in Table~2, will appear only in the electronic edition. The table
contains x and y positions of each star on the chip, CL and LP magnitudes
with their respective uncertainties as well V and I magnitudes calibrated 
using the calibration equations (1)--(3). The uncertainties given 
for V and I magnitudes
come from the errors in CL and LP photometry and from the transformations.

Our calibration equations are only preliminary and valid for a limited
color range ($-0.8<(V-I)<2$). Since essentially all our comparison stars 
are red giants or supergiants, the transformation may be different for dwarfs 
(Gregg \& Minniti 1997). The complete calibration of the STIS photometry to 
the standard photometric system, as well as a more quantitative analysis of the
influence of different physical parameters (i.e. temperature, abundances, 
surface gravity) on the transformation, 
will be published elsewhere (Goudfrooij et al. in preparation).

\section{Fundamental parameters of WLM}

Figure~6 shows the final calibrated $V ~vs ~V-I$ and $I ~vs ~V-I$
color-magnitude diagrams. These diagrams
are typical for a disk + halo population in WLM. 
The main features are a red giant branch (RGB),
a blue main sequence with mean color $V-I = -0.2$, a prominent red 
giant clump at $I=24.68\pm 0.05$ and  $V-I=0.9$, and a 
horizontal branch (see also Fig.~4). 
The horizontal branch (HB) is extended, and can be traced to
$V \simeq 26$, and $V-I \simeq 0$. 

Using the deep color-magnitude diagram of Figure~6 we can 
redetermine fundamental
WLM parameters such as distance, metallicity and age. It is beyond the scope
of this paper to determine the detailed star formation history in the field.

\subsection{The Distance}

Recent Hipparcos observations accurately measured the absolute magnitudes
of the old and metal-poor blue HB stars of the Milky Way halo.
For these stars, Hipparcos gives $M_V=0.7$
for [Fe/H]$=-1.5$ (de Boer 1999). This metallicity is 
typical for the WLM halo, allowing one to obtain an accurate distance 
to this galaxy by comparing directly 
the V-band magnitude of the HB. 

We created V-band luminosity functions and derivatives, with different 
magnitude bins and with different bin starting points, for stars with color
$0.0<V-I<0.55$, in order to exclude contamination by the red giant clump.
In this color range there is still small contamination by the reddest stars
of the main sequence. We tried also other different color ranges and 
different binning procedures, but the results were unchanged. 
The V-band magnitudes of the HB, calculated from individual luminosity 
function derivatives, were averaged to obtain the mean V magnitude of 
the HB: $m_V^{HB}=25.71\pm0.09$. Two examples of typical 
luminosity functions 
created for a bin size of 0.2 mag are given in Figure~7. A luminosity
function for the color range $0.0<V-I<0.55$ is plotted with the solid line, 
while the dotted line is used for the one covering $0.15<V-I<0.55$.
The arrow indicates the average V magnitude of the blue HB (which is a
$3\sigma$ detection here) obtained from many different luminosity function
derivatives. The strength of the HB
feature in the luminosity function shows that it is indeed
extended beyond the red giant clump. In such an extended HB we expect to
have some RR Lyrae stars which can widen the feature in the CMD, although
unfortunately we do not have temporal sampling to find them.

If we adopt $E_{V-I}=0.03$ from Minniti \& Zijlstra (1997), 
the distance modulus turns out to be $(m-M)_0=24.95\pm0.09$, including only
our statistical error. However,
the final conservative distance error we adopt is $\sigma=0.13$, in order to
account for any systematic error in the Hipparcos results.
Using the magnitude of the tip of the red giant branch (RGBT), 
Lee et al.\ (1993) measured a
distance modulus of $(m-M)_0=24.81\pm0.15$, 
and Minniti \& Zijlstra (1997) measured $(m-M)_0=24.75\pm0.10$.
The Cepheid distance modulus measured by 
Lee et al.\ (1993), based on a distance modulus 
to LMC of $(m-M)_0^{LMC} = 18.5$, is $(m-M)_0=24.92\pm0.21$. 
The present distance modulus to WLM measured using the HB stars 
blueward of the red giant clump, 
$(m-M)_0 = 24.95$, is within $1\sigma$ of the former and $2\sigma$ 
of the latter RGBT distances mentioned above, 
and is very close to the Cepheid distance. 
It is also in good agreement with $(m-M)_0=24.88\pm0.09$, the 
distance modulus determined from WFPC HST data by fitting the entire
color-magnitude diagram (Dolphin 2000). 
Therefore, we conclude that for $E_{V-I}=0.03$, the
distance modulus to WLM is $(m-M)_0 = 24.95 \pm 0.13$. This translates to a
distance of $D = 0.98\pm0.06$ Mpc, within $1\sigma$ of
$D_{WLM} = 0.94$ Mpc listed by van den Bergh (1994).

\subsection{Metal Abundances}

The metallicity of the old and metal-poor populations such as those
found in the halo around WLM, can be derived from the V$-$I color of
their RGB.  Using the calibration of Lee (1993), Minniti \& Zijlstra
(1997) obtained [Fe/H]$\, = -1.45\pm 0.2$. The full line on Figure~8 
compares the location of the RGB, AGB and HB of the globular cluster M5
(Johnson \& Bolte 1998) with the WLM photometry. The
globular cluster fiducial lines were brought to the distance of WLM (see
Sec.~3.1) adopting the distance modulus of 14.35 and the reddening of 
E(B-V)=0.03 for M5.
The metallicity of M5 in Zinn \& West (1984) is 
[Fe/H]$_{M5}\,=-1.40$. The mean
color of the RGB from the STIS color-magnitude diagrams is similar to
that obtained with the NTT data, and to the fiducial RGB sequence of
M5, therefore confirming the metallicity determined by Minniti \&
Zijlstra (1997). 

The RGB in our CMD appears wider than that of Dolphin's (2000)
CMDs. These are different fields, and the effect can be due to a) different
metallicities, b) different ages, c) bigger errors in the STIS V-I colors,
d) differential reddening, or e) a combination of the above.
The CMDs clearly show the presence of young main sequence stars as well
as old horizontal branch stars in this field. For comparison, we show the 
effects of old populations of different metallicities in Figure~8, where
the dashed lines are the RGB fiducials for
Galactic globular clusters M15, NGC 6752 and 47 Tuc (Da Costa \& Armandroff
1990), from left to right respectively. These globular clusters span a wide
range in 
metallicities, from $-2.17$ dex to $-0.71$ dex. Note that there are
stars more metal poor than the mean metallicity, as well as the absence of
metal-rich stars.

By fitting the theoretical isochrones Hodge et al.\
(1999) find a metallicity of [Fe/H]$\, = -1.52\pm 0.08$ for the only
globular cluster detected in WLM, in good agreement with the
metallicity obtained for the outer field giant stars by Minniti \&
Zijlstra (1997).

\subsection{The Ages}

Blue HB stars, like RR Lyrae stars, are unambiguous 
indicators of old populations with 
age $>10^{10}$ yr (Da Costa 1994;
Hodge 1989; Olszewski et al. 1994; Grebel 1998; Mateo 1998).
The detection of a blue horizontal branch in WLM confirms the presence
of ancient stars. Based on the HB morphology, we predict 
the existence of RR Lyrae variables in this galaxy. 

Figure~9 shows the photometry of the field 
compared with the most recent Padova isochrones.
There is a good agreement with the isochrones of 
Girardi et al. (2000) for $Z=0.001$, $Y=0.23$, and $\log~t=10.20$ and 
$7.80$. These indicate a mixture of an old and young population,
characteristic of the halo and disk respectively, as expected for a field
situated at the very edge of the WLM disk.

By comparing with
theoretical isochrones and luminosity functions Sandquist et al.\ (1996)
measure an absolute age for M5 of $13.5\pm1$ Gyr or $11\pm1$ Gyr, for
[Fe/H]$_{M5}=-1.40$ or [Fe/H]$_{M5}=-1.17$ respectively. 
The good
agreement of our photometry and the fiducial sequence of M5 (Figure~8)
is another
strong argument in favor of the presence of an old stellar population with
age$>10^{10}$ yr in the halo of WLM.

Hodge et al.\ (1999) obtain an absolute age of $14.8\pm0.6$ Gyr for the 
the only globular cluster detected in WLM. 
The age and metallicity of this globular cluster
is similar to normal old halo globulars in our Galaxy. This confirms
again the presence of an old and metal-poor population in WLM similar to
the one in the Milky Way halo and to the populations found in
the halos of the other spirals of the Local Group (Mould \& Kristian 1986).

Following its early phase of star formation, a galaxy like WLM could
have been accreted to form the Milky Way halo. It would have given its
stellar content to the halo, which today we observe to be of similar
age and metallicity as the oldest stars in WLM (see also the
comparison between WLM and M5 Galactic globular cluster in
Figure~8). The remaining gas from a WLM-like galaxy would have sunk
down to the bottom of the potential well and contribute to the gas
used to form stars in the disk of the Milky Way.  Because of this, WLM
can be considered as an example of a surviving protogalactic fragment,
related to  those that built up the Milky Way halo (Minniti \& Zijlstra
1996).

\section{Conclusions and Implications}

We have obtained deep STIS $CL$ and $LP$-band photometry for 1886
stars in the field of the dwarf
irregular galaxy WLM, located at the edge of the Local Group. 

We have calibrated the STIS photometric system empirically onto the
$V$ and $I$-band system using our previous observations with the NTT 
telescope (Minniti \& Zijlstra 1997).
The VI color-magnitude diagram of a field located in the transition region
between WLM's disk and halo shows a prominent RGB, a blue main sequence, 
a red clump,
and an extended HB. The presented CMD was obtained with HST+STIS over less
than 2 orbits, and revealed the presence of a HB in this distant Local Group
galaxy. These results prove that it is now possible to reach the 
blue horizontal branch out to $D\simeq 1$ Mpc using STIS with relatively
short exposure times ($\la 1$ orbit), allowing
the study of the horizontal branches throughout the Local Group.

We have obtained an accurate distance modulus for WLM of
$(m-M)_0=24.95\pm 0.13$, based on the magnitude of the horizontal 
branch. This value agrees well with recent
distance determinations using the RGB tip and Cepheids.

Our main result is, however, the detection of a horizontal branch that
extends blueward from the red giant clump,
an unambiguous indicator of the presence of an old population
(age$ > 10^{10}$ yr) in WLM. It is likely to be associated with the halo of 
the galaxy, which is found to be an  extended metal-poor component by
Minniti \& Zijlstra (1996). The old age measured for the globular cluster 
of this galaxy by Hodge et al. (1999) favors such an interpretation.

Kinematics of individual stars is needed to compare the disk population with
the halo population to check if the latter is kinematically hotter (with
higher velocity dispersion and lower rotational velocity).

There are two major implications of these results:

1) WLM formed stars more than 10 Gyr ago. Various cosmological models give
different redshifts for the same lookback time (Carroll, Press \& Turner
1992). For example, according to the model of a flat universe with a 
plausible matter ($\Omega_M=0.1$) and without the cosmological constant 
($\Omega_{\Lambda}=0$), WLM formed stars at $z>4$, while according to the 
cosmological model of a flat universe with $\Omega_M=0.1$ and 
$\Omega_{\Lambda}=0.9$, WLM formed stars at $z>1$.
Star formation in this galaxy was not inhibited
at early times by photoionization by UV sources such as QSOs and AGNs. 
This does not favor scenarios where
dwarf galaxies form recently (e.g. Babul \& Rees 1992).

2) The old population found in the WLM halo is similar to that of the old
and metal-poor halos of the spiral galaxies in the Local Group. Having
remained isolated throughout its lifetime, 
WLM can now be considered as a good example of a surviving 
protogalactic fragment, related to those that built up the Milky Way halo. 

\acknowledgements{
DM is grateful to his colleagues of
the Parallel Working Group at STScI who made these observations possible.
This work was performed in part under the auspices of the Chilean Fondecyt
No.\ 01990440, and 07990048, and of
DIPUC No.\ 98.16E, and by the U. S. Department of 
Energy by Lawrence Livermore National Laboratory under Contract W-7405-Eng-48. }
MVA acknowledges financial support from the SECYT and CONICET. 	
We are grateful to George Hau for many useful comments and suggestions.

\clearpage
\begin{deluxetable}{rrrrrrrll}
\footnotesize
\tablenum{1}
\tablecaption{NTT VI vs. STIS CL-LP photometry: positions and magnitudes of
the 41 stars in
common in both the NTT and STIS fields. \label{tbl-1}}
\tablewidth{0pt}          
\tablehead{
\colhead{\#} & \colhead{$x(NTT)$} & \colhead{$y(NTT)$}
& \colhead{$V$} & \colhead{$I$} & \colhead{$x(STIS)$} & \colhead{$y(STIS)$}
& \colhead{$CL$} & \colhead{$LP$}
}
\startdata
 1&816.9&1937.9&22.59&21.51&81.998& 169.736&22.71&22.65\nl
 2&803.1&1922.3&23.16&21.69&96.888& 59.512& 23.17&22.93\nl
 3&829.3&1908.6&22.97&21.44&241.726&118.313&22.75&22.59\nl
 4&823.5&1887.3&23.68&22.59&305.413&21.158& 24.21&24.19$^{**}$\nl
 5&890.6&1941.2&24.12&22.65&322.588&475.961&23.67&23.55\nl
 6&901.8&1944.4&21.95&21.76&347.604&532.426&22.26&22.61\nl
 7&892.4&1921.4&23.36&22.42&406.552&414.659&23.53&23.59\nl
 8&883.7&1906.3&23.40&22.26&438.210&329.957&23.80&23.75\nl
 9&867.5&1890.2&22.32&21.23&446.591&208.103&22.27&22.25\nl
10&910.5&1920.0&21.25&21.16&474.453&476.121&24.84&25.62$^{**}$\nl
11&884.3&1897.6&23.35&21.78&474.557&301.366&23.02&22.89\nl
12&847.5&1864.3&22.99&21.67&481.419&39.644& 22.99&22.94\nl
13&914.7&1915.4&21.53&21.54&509.622&484.332&21.95&22.49$^{**}$\nl
14&894.2&1893.9&23.81&23.03&523.888&326.995&23.14&23.61$^{**}$\nl
15&909.5&1902.4&23.75&22.68&542.293&418.119&23.30&23.44\nl
16&867.5&1861.7&22.80&22.00&561.449&109.459&22.75&22.98\nl
17&880.4&1864.7&23.60&22.43&592.676&170.379&23.88&23.83\nl
18&877.4&1862.0&22.92&20.82&593.314&150.726&22.37&22.12\nl
19&910.9&1888.1&25.14&23.03&605.225&377.214&24.44&24.44$^{**}$\nl
20&878.5&1850.4&21.76&21.50&642.896&115.016&21.89&22.41\nl
21&874.8&1845.3&21.41&19.95&651.276&82.676& 21.25&21.13\nl
22&882.0&1845.2&23.08&21.62&676.151&111.879&22.97&22.83\nl
23&931.0&1890.3&23.89&22.94&687.633&452.606&24.01&23.99\nl
24&919.5&1874.3&22.36&20.91&688.662&361.977&22.35&22.19\nl
25&876.0&1835.1&24.25&22.80&696.083&53.898& 23.64&23.71$^{*}$\nl
26&910.7&1856.6&23.71&22.38&729.336&266.375&23.41&23.38\nl
27&885.0&1826.5&22.00&20.77&761.784&59.269& 21.99&21.96\nl
28&889.1&1820.5&22.33&21.11&799.237&55.663& 22.39&22.34\nl
29&892.2&1811.9&23.71&22.72&847.889&38.486& 23.95&24.16\nl
30&940.1&1850.3&23.25&22.04&855.515&362.924&23.13&23.16\nl
31&908.3&1822.4&22.90&22.99&857.748&140.250&22.78&23.67\nl
32&910.5&1816.7&23.10&22.01&889.823&127.514&23.22&23.18\nl
33&914.0&1813.2&23.73&23.20&910.690&126.506&24.38&24.79$^{**}$\nl
34&943.6&1838.4&23.13&21.83&914.359&331.962&23.60&23.52\nl
35&971.3&1855.4&22.96&22.10&943.155&503.628&23.16&23.27\nl
36&921.7&1811.2&22.74&21.77&948.750&153.724&22.74&22.81\nl
37&965.9&1847.5&21.56&20.84&956.147&455.374&21.92&22.11\nl
38&977.9&1857.3&22.48&21.14&958.390&536.273&22.55&22.53\nl
39&917.1&1802.4&22.45&20.99&968.605&105.047&22.21&22.11\nl
40&928.8&1812.6&23.74&24.24&969.604&182.020&23.66&24.42$^{*}$\nl
41&937.8&1819.4&23.28&22.35&971.038&247.251&23.35&23.43\nl

\enddata


\tablenotetext{*}{marginal detection in the NTT image}
\tablenotetext{**}{multiple sources in the STIS image}

\end{deluxetable}

\clearpage

\begin{deluxetable}{cccccccccc}
\footnotesize
\tablenum{2}
\tablecaption{The complete photometry list.$^{*}$
}
\tablewidth{0pt}
\tablehead{
\colhead{x} & \colhead{y} & \colhead{CL}
& \colhead{$\sigma$(CL)} & \colhead{LP} & \colhead{$\sigma$(LP)} 
& \colhead{V}
& \colhead{$\sigma$(V)} & \colhead{I} & \colhead{$\sigma$(I)}
}
\startdata
1.338&376.277&26.594&0.054&27.400&0.106&26.40&0.19&26.64&0.21\nl
5.169&97.329&27.412&0.092&27.479&0.143&27.41&0.14&26.35&0.24\nl
6.707&504.232&26.415&0.059&26.428&0.047&26.42&0.08&25.27&0.10\nl
7.505&500.302&27.658&0.109&28.046&0.156&27.57&0.19&27.08&0.27\nl
8.480&553.033&27.345&0.099&27.189&0.123&27.40&0.15&25.94&0.22\nl

\enddata


\tablenotetext{*}{The complete version of this table is in the electronic
edition of the Journal.  The printed edition contains only a sample.}

\end{deluxetable}

\clearpage
\begin{figure}
\caption{
NTT V-band image of WLM, covering $1\times 1$ arcmin$^2$.
North is up and East is to the left. At the distance of WLM,
$1\arcmin = 0.26 ~kpc$. The point sources in this image have
FWHM$=0.8\arcsec$.
The box indicates the location of the $50\arcsec \times 28\arcsec$ STIS
field.}
\end{figure}

\begin{figure}
\caption{
STIS image of the WLM field with the $LP$ filter, covering 
$50\arcsec \times 28\arcsec$.
The point sources in this image have FWHM$=0.1\arcsec$.
}
\end{figure}

\begin{figure}
\scalebox{0.62}{\rotatebox{270}{\includegraphics{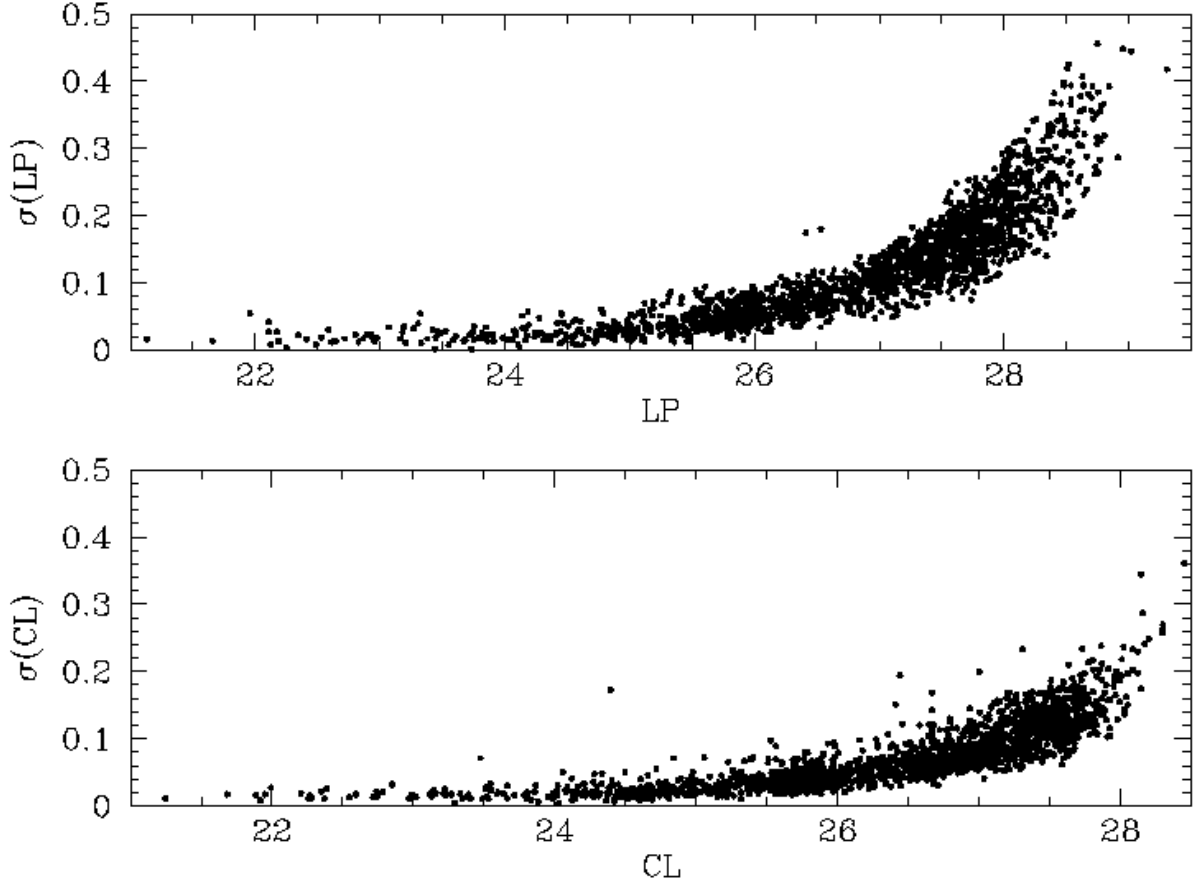}}}
\caption{
Photometric errors as a function of instrumental
$LP$ (top) and $CL$-band (bottom) magnitudes.
}
\end{figure}

\begin{figure}
\scalebox{0.62}{\rotatebox{270}{\includegraphics{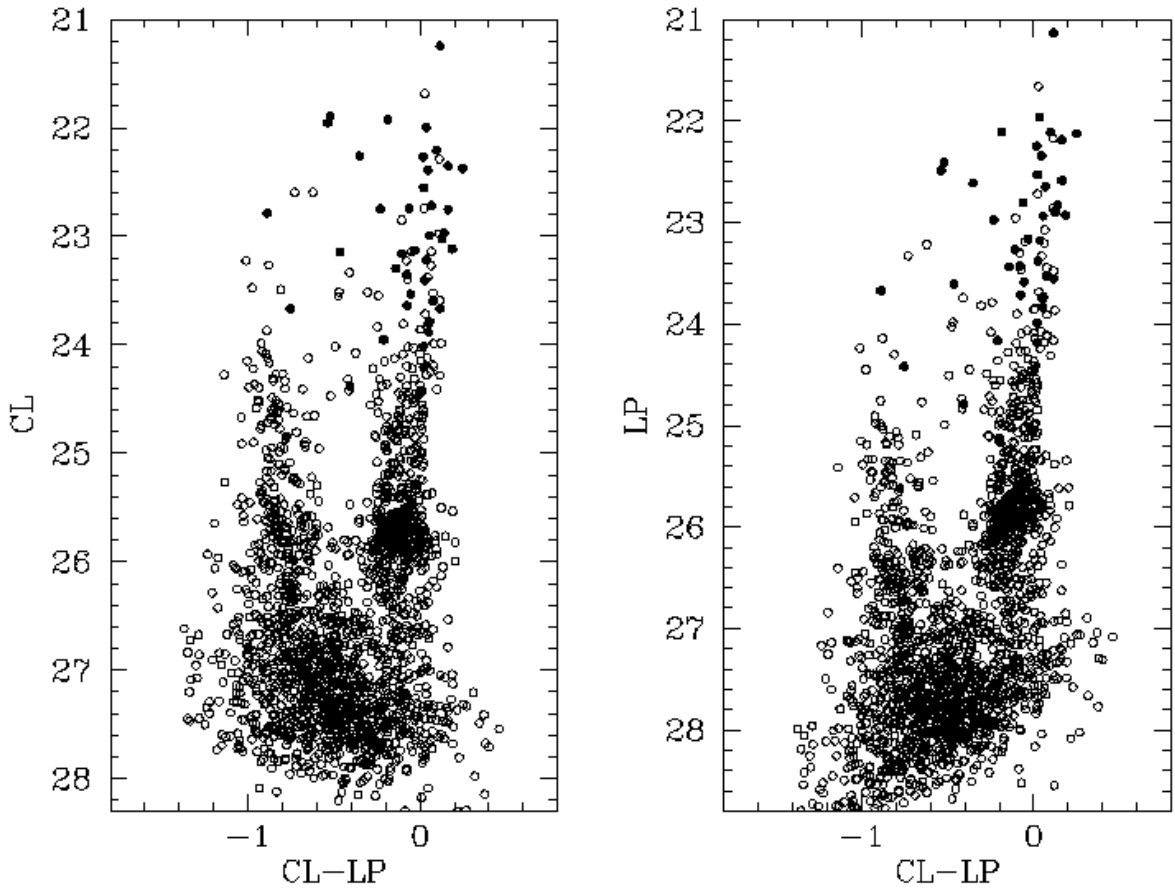}}}
\caption{
Color-magnitude diagrams of the WLM field in the STIS photometric system.
The filled circles show the stars in common with the NTT data, which are 
used to calibrate the STIS photometric system.
}
\end{figure}

\begin{figure}
\scalebox{0.62}{\rotatebox{270}{\includegraphics{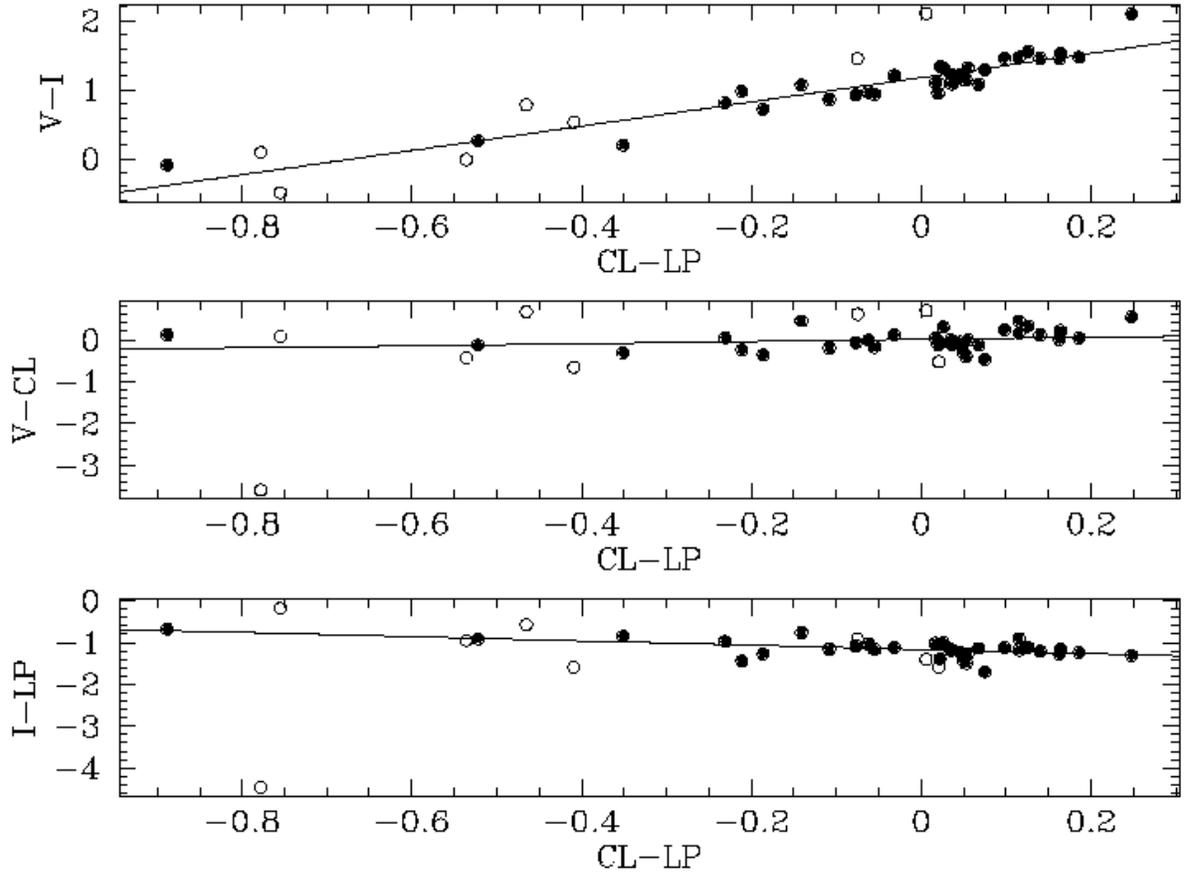}}}
\caption{
Empirical calibration of the STIS photometric system 
($CL$, $LP$ and $SP$-bands) into the V and I-bands. Only the 33 stars
(filled circles) out of the total 41 (open and filled circles), found in
both NTT and STIS fields, were used for the calibration.
}
\end{figure}

\begin{figure}
\scalebox{0.62}{\rotatebox{270}{\includegraphics{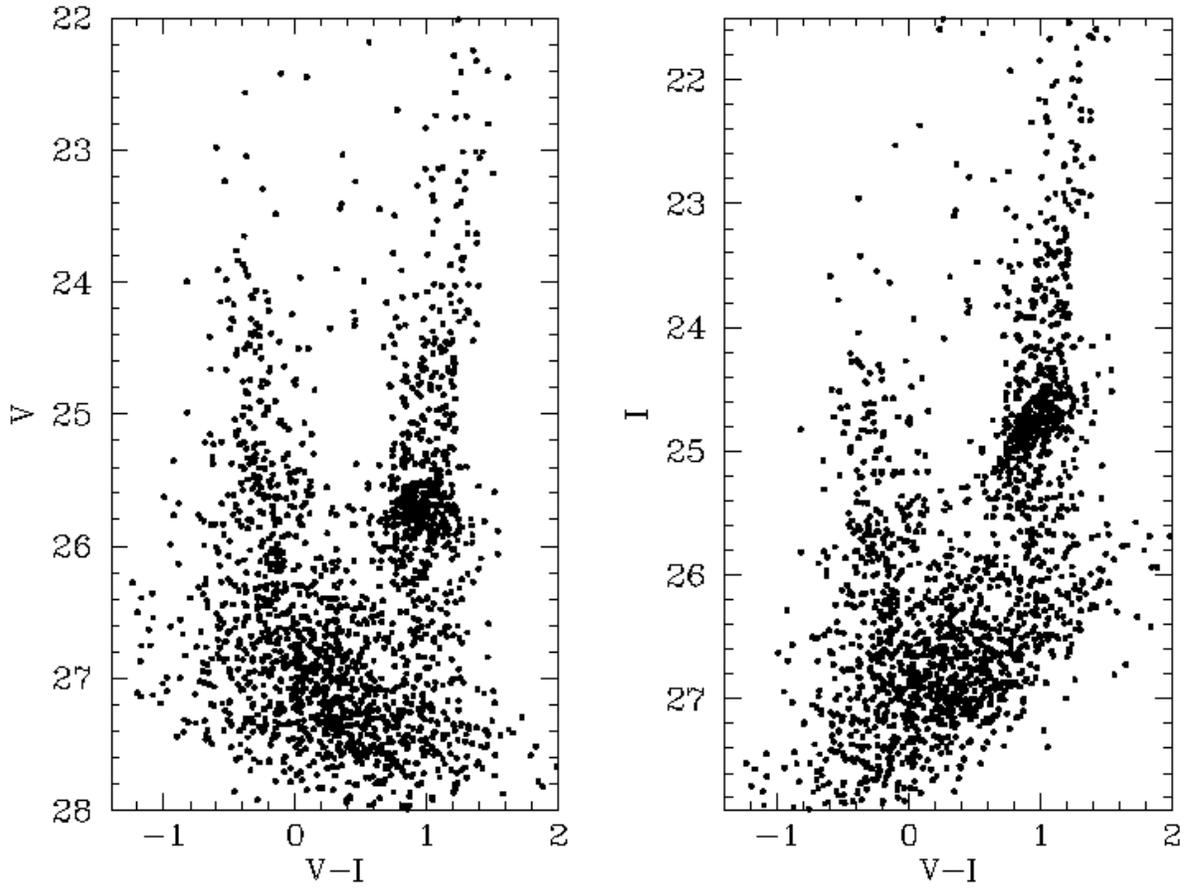}}}
\caption{
Final calibrated $VI$ color-magnitude diagrams of the WLM STIS field.
}
\end{figure}

\begin{figure}
\scalebox{0.62}{\rotatebox{270}{\includegraphics{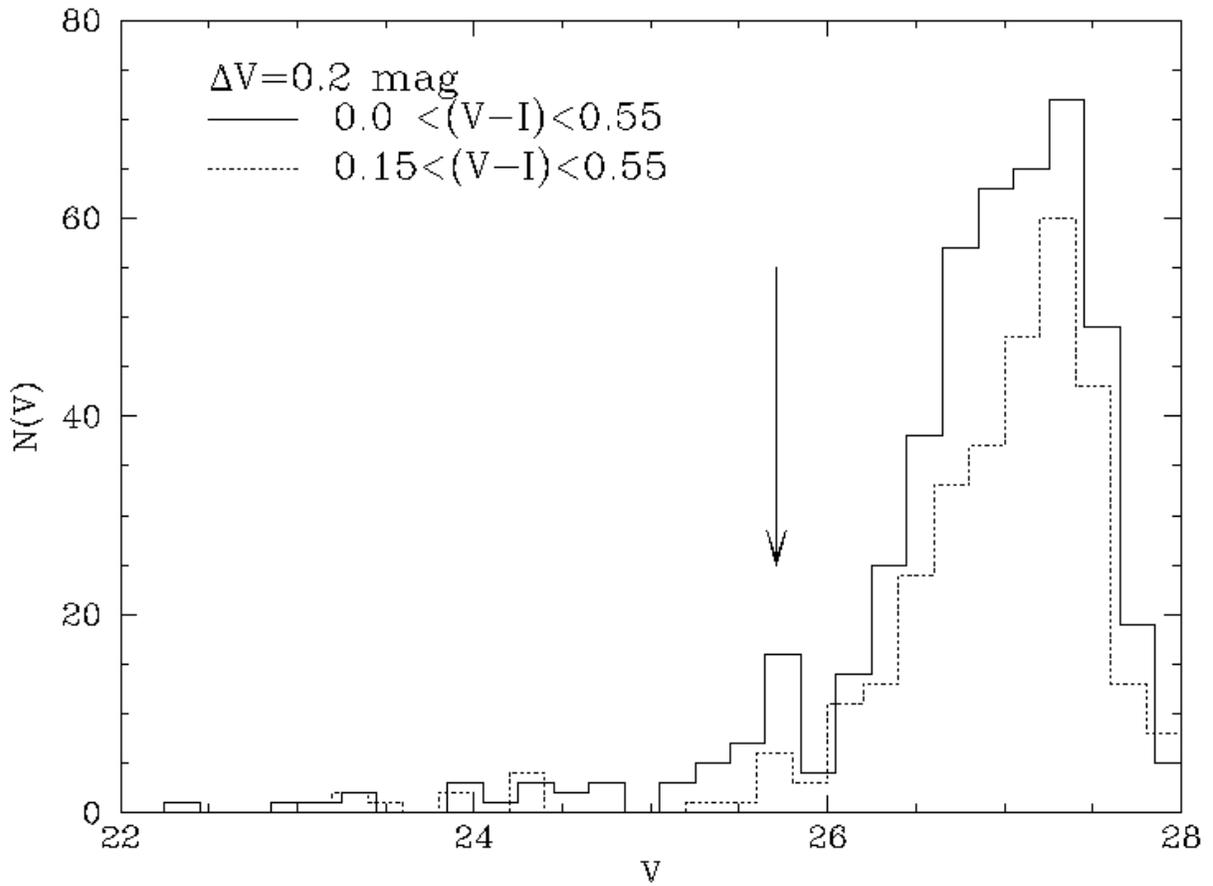}}}
\caption{
Two typical V-band luminosity functions for the magnitude bin of 0.2 mag
and two different color ranges, 0.00$<$V$-$I$<$0.55 (solid line) and
0.15$<$V$-$I$<$0.55 (dotted line).
The arrow indicates the position of the horizontal branch. }
\end{figure}

\begin{figure}
\scalebox{0.62}{\rotatebox{270}{\includegraphics{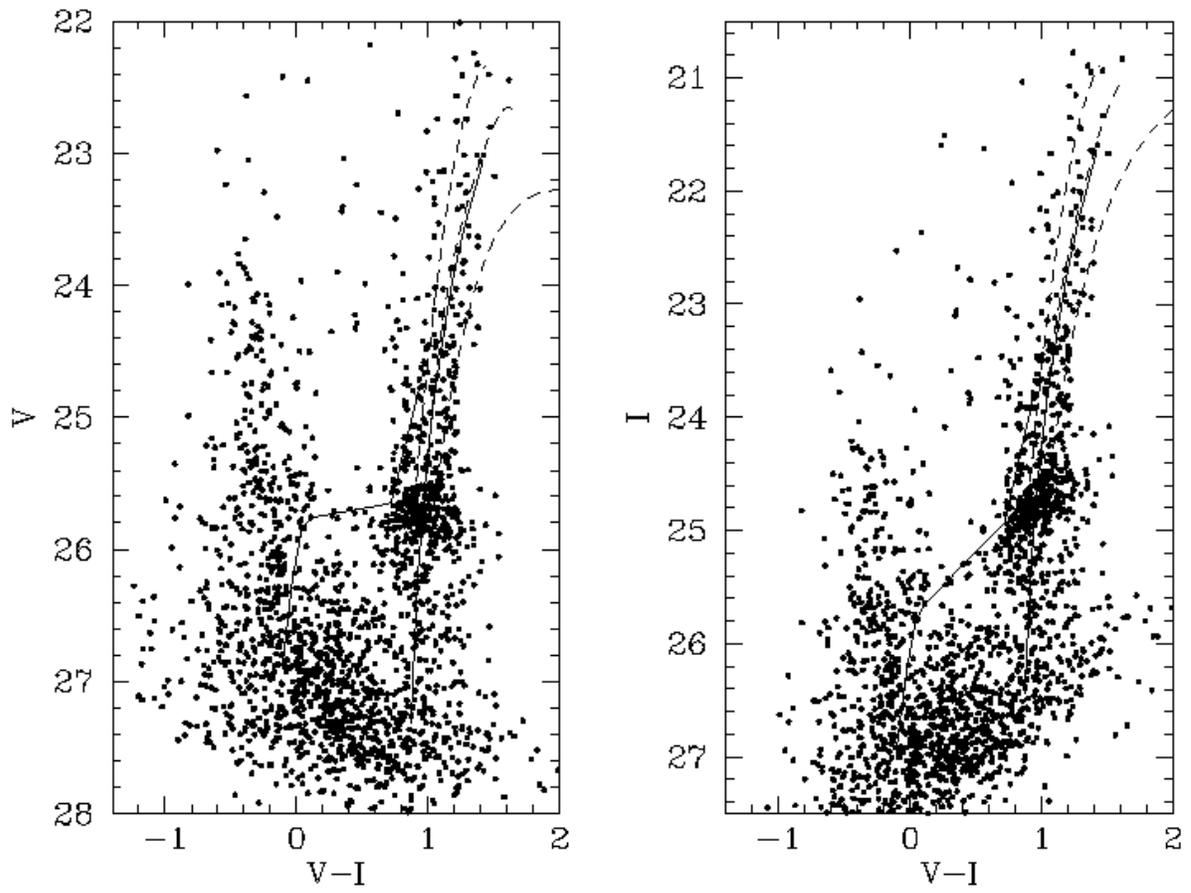}}}
\caption{
Comparison of the WLM $VI$ color-magnitude diagram with that of the galactic
globular clusters M5 (solid line; Johnson \& Bolte 1998) and the RGBs of M15,
NGC 6752 and 47 Tuc (dashed lines; Da Costa \& Armandroff 1990). 
}
\end{figure}

\begin{figure}
\scalebox{0.62}{\rotatebox{270}{\includegraphics{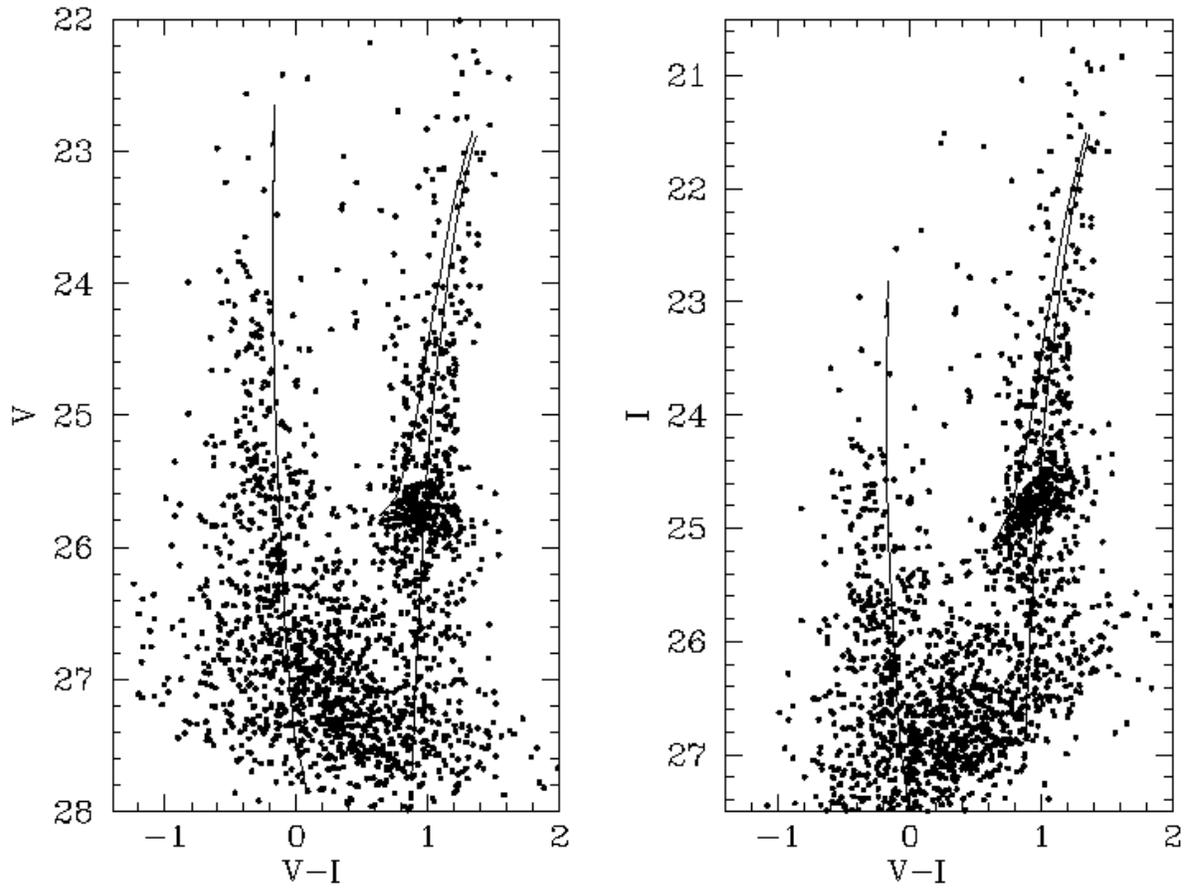}}}
\caption{
Comparison of the WLM $VI$ color-magnitude diagram with the Padova 
isochrones (Girardi et al.\ 2000). Isochrones for
$Z=0.001$ and $log\, t=7.80$ and $10.20$ are plotted over the data. }
\end{figure}

\end{document}